\documentclass[
reprint,
superscriptaddress,
 amsmath,amssymb,
 aps,
]{revtex4-2}

\usepackage{graphicx}
\usepackage{bm}
\usepackage[colorlinks=true,citecolor=red,linkcolor=blue,urlcolor=blue]{hyperref}

\begin{document}

\title{Chirotactic response of microswimmers in fluids with odd viscosity}

\author{Yuto Hosaka}
\email{yuto.hosaka@ds.mpg.de}
\affiliation{Max Planck Institute for Dynamics and Self-Organization (MPI-DS), Am Fassberg 17, 37077 G\"{o}ttingen, Germany}

\author{Michalis Chatzittofi}
\affiliation{Max Planck Institute for Dynamics and Self-Organization (MPI-DS), Am Fassberg 17, 37077 G\"{o}ttingen, Germany}

\author{Ramin Golestanian} 
\email{ramin.golestanian@ds.mpg.de}
\affiliation{Max Planck Institute for Dynamics and Self-Organization (MPI-DS), Am Fassberg 17, 37077 G\"{o}ttingen, Germany}
\affiliation{Rudolf Peierls Centre for Theoretical Physics, University of Oxford, Oxford OX1 3PU, United Kingdom}

\author{Andrej Vilfan} 
\email{andrej.vilfan@ds.mpg.de}
\affiliation{Max Planck Institute for Dynamics and Self-Organization (MPI-DS), Am Fassberg 17, 37077 G\"{o}ttingen, Germany}
\affiliation{Jo\v{z}ef Stefan Institute, 1000 Ljubljana, Slovenia}

\begin{abstract}
Odd viscosity is a property of chiral active fluids with broken time-reversal and parity symmetries. We show that the flow of such a fluid around a rotating axisymmetric body is exactly solvable and use this solution to determine the orientational dynamics of surface-driven microswimmers. Swimmers with a force-dipole moment exhibit precession around the axis of the odd viscosity. In addition, pushers show bimodal chirotaxis, i.e., alignment parallel or antiparallel to the axis, while pullers orbit in a plane perpendicular to it. A chiral swimmer that itself has a broken parity symmetry can exhibit unimodal chirotaxis and always align in the same direction.
\end{abstract}

\maketitle

The diversity and versatility of nonequilibrium behavior in active matter systems~\cite{gompper2020,marchetti2013hydrodynamics,cates2015motility,Golestanian2018phoretic} and the possibilities to control them have in recent years provided a promising roadmap to the realization of new forms of self-organization~\cite{Snezhko2011,Zhou2014,Guillamat2016,Cohen2014,Ellis2017,Waisbord2016,Aubret2018,Vincenti2018,MassanaCid2019,Matsunaga2019,Han2020}. To understand such complex phenomena, it is essential to be able to perform a systematic bottom-up characterization of the mechanisms of nonequilibrium activity and the responses of individual active agents to external cues as well as physical interactions from other agents. Such systematic coarse-graining approaches can reveal nontrivial large-scale collective phenomena with emergent properties, which cannot always be predicted from general symmetry considerations. In living systems, microorganisms adapt their movement to environmental cues through a series of processes called taxes \cite{Bray2000}. These include chemotaxis~\cite{eisenbach2004chemotaxis} and thermotaxis~\cite{Bahat.Eisenbach2003}, as well as alignment responses such as phototaxis~\cite{Jekely2009,bennett2015steering,RayPhoto2023}, magnetotaxis~\cite{blakemore1975magnetotactic}, gravitaxis or the related gyrotaxis for bottom-heavy organisms~\cite{kesslerIndividual}, and rheotaxis in a shear flows~\cite{Marcos.Stocker2012}. Similar responses have also been realized in various artificial microswimmers~\cite{ten2014gravitaxis,Das2015,Simmchen2016,Lozano.Bechinger2016,Jin.Maass2017}.

Although microswimmers, whether living or synthetic, have been studied in aqueous environments, they themselves can exhibit active properties by violating symmetries that would be conserved in classical systems~\cite{shankar2022topological}. Fluids with broken time-reversal and parity symmetries can have a peculiar transport coefficient called \textit{odd viscosity}~\cite{avron1995, avron1998, banerjee2017}. This viscosity has a number of unusual properties~\cite{hosaka2022nonequilibrium,fruchart2023odd}, such as being dissipationless, allowing topological sound waves~\cite{souslov2019topological, tauber2019} and leading to nonreciprocal (transverse) response. The latter appears as the antisymmetric component in the Green function~\cite{hosaka2021nonreciprocal, khain2022, hosaka2023pair, yuan2023stokesian, lier2023lift} or in the resistance tensor (or its consequence as lift or Magnus forces)~\cite{hosaka2021hydrodynamic, lier2023lift, hosaka2023lorentz}. The asymmetric resistance tensor in such chiral active flows indicates the violation of the Onsager reciprocity that requires a conserved time-reversal symmetry~\cite{doi2013soft, doi2015onsager}. In artificial or even living systems, chiral active matter is abundant on nano- to macroscales, e.g., from rotary membrane proteins~\cite{oppenheimer2019} to active nematic cells~\cite{Thampi2016,yamauchi2020, PhysRevX.12.041017} or chiral assemblies of bacteria \cite{uchida2010synchronization, uchida2010synchronization2}, colloidal spinners~\cite{ebbens2010,MassanaCid2019,Matsunaga2019,massana2021arrested, bililign2022motile,vilfan2023spontaneous}, swimming algae~\cite{drescher2009dancing} or starfish embryos~\cite{tan2022odd}. In particular, odd viscosity has been observed in a fluid of rotating particles in experimental~\cite{soni2019odd, yang2021topologically,mecke2023simultaneous} and numerical studies~\cite{hargus2020time,han2021fluctuating,zhao2021,lou2022odd}.
When a microswimmer, either area-changing~\cite{lapa2014}, dipolar~\cite{hosaka2023hydrodynamics}, or surface-driven~\cite{hosaka2023lorentz}, is immersed in a fluid with odd viscosity, it can exhibit transverse motion.
This leads to the question about the orientational dynamics of a microswimmer in an odd-viscous fluid.

\begin{figure}[b]
\centering
\includegraphics[width=0.83\columnwidth]{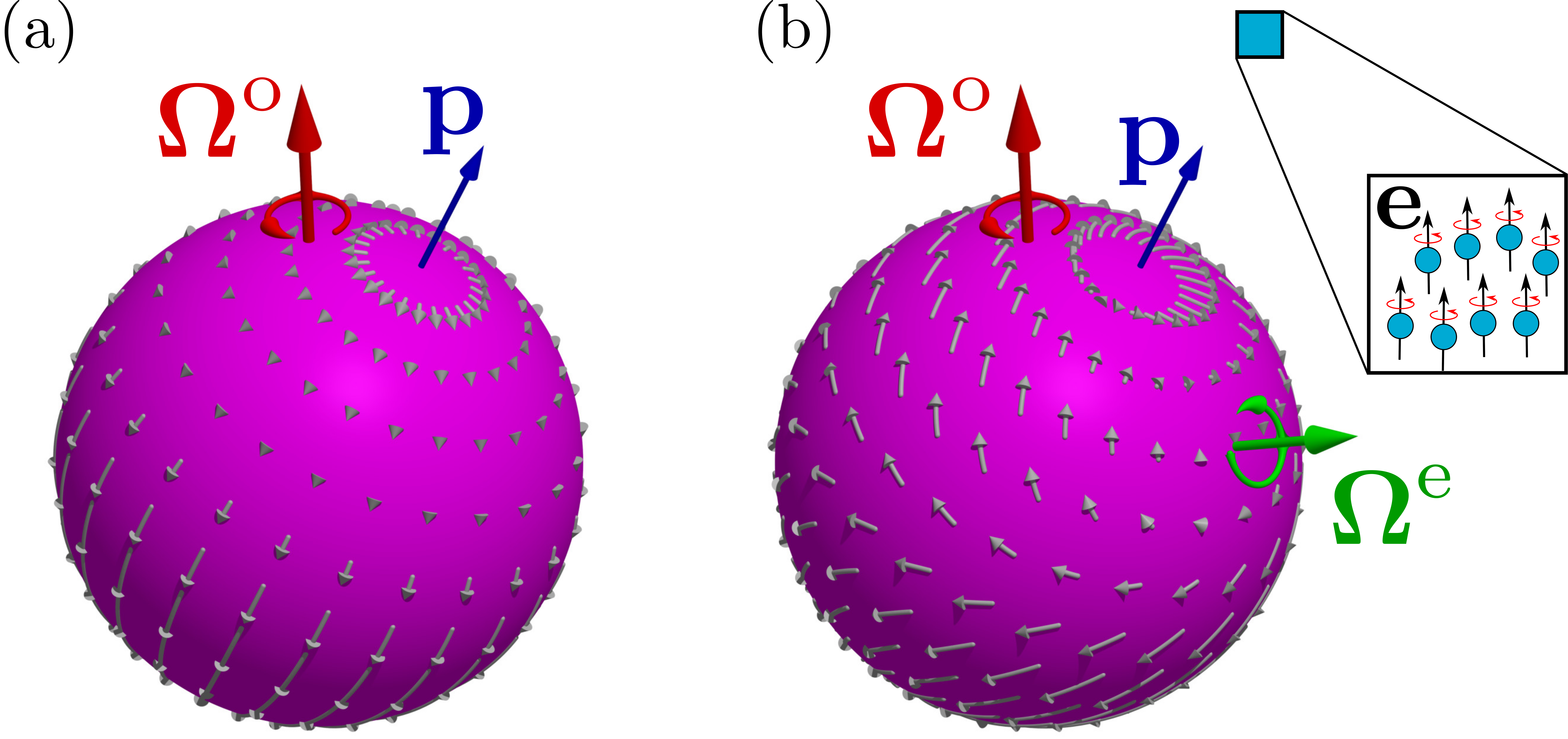}
\caption{
A surface-driven microswimmer with orientation $\mathbf{p}$ immersed in a three-dimensional fluid with an odd viscosity $\eta^{\rm o}$ (e.g., a fluid composed of the microscopic components that collectively self-rotate around the axis $\mathbf{e}$).
(a) An achiral pusher shows rotation with angular velocity $\boldsymbol{\Omega}^{\rm o}$, which leads to alignment in $\pm\mathbf{e}$-directions (\textit{bimodal chirotaxis}).
(b) A chiral swimmer with an additional intrinsic angular velocity $\mathbf{\Omega}^{\rm e}$ can align in a single direction (\textit{unimodal chirotaxis}).
\label{fig:1}
}
\end{figure}
\begin{figure}[tb]
\centering
\includegraphics[width=0.96\columnwidth]{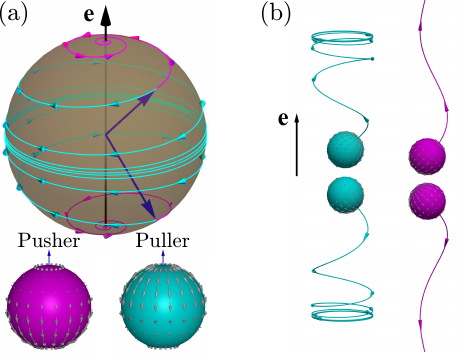}
\caption{Bimodal chirotaxis. 
(a) Time evolution of the swimmer orientation $\mathbf{p}$ with the squirming mode $B_2$ [Eq.~(\ref{eq:pdot})] with two different initial orientations (blue arrows). 
Pullers ($B_2>0$, cyan) converge towards circling in a plane perpendicular to $\mathbf{e}$. Pushers ($B_2<0$, magenta) align towards $+\mathbf{e}$ or $-\mathbf{e}$ directions, depending on their initial orientation.
(b) Trajectories followed by pullers and pushers of different initial orientations. All trajectories are computed with the odd-to-even viscosity ratio of $\lambda=0.3$.
\label{fig:2}
}
\end{figure}

%
Here we discuss the rotational dynamics of surface-driven spherical microswimmers (squirmer) in fluids with odd viscosity (Fig.~\ref{fig:1}).
By finding an exact solution of the ``odd'' Stokes flow around a rotating axisymmetric body and applying the generalized Lorentz reciprocal theorem, we derive the angular velocity of a swimmer with an arbitrary surface velocity profile.
We show that the force-dipole (stresslet) moment of a swimmer interacting with the chiral nature of the fluid can lead to intriguing orientational dynamics (Fig.~\ref{fig:2}), unlike standard fluids without odd viscosity.
In analogy to the tactic response of microorganisms, we denote this effect as \textit{chirotaxis}.
In fluids with odd viscosity (axis of chirality $\mathbf{e}$), the angular velocity of a swimmer with orientation $\mathbf{p}$ is given by
\begin{align}
\boldsymbol{\Omega}^{\rm o}= 
\Omega_{\rm Sp}(\mathbf{e}\cdot\mathbf{p})\mathbf{p} 
+ \Omega_{\rm Pr}\mathbf{e}
+ \Omega_{\rm Ch} (\mathbf{e}\cdot\mathbf{p})\mathbf{e}\times\mathbf{p}
\,.
\label{eq:intro}
\end{align}
While undergoing spinning [$\Omega_{\rm Sp}$; see Eq.~(\ref{eq:Sp})] and precession [$\Omega_{\rm Pr}$; see Eq.~(\ref{eq:Pr})] around the axis $\mathbf{e}$, the swimmer exhibits \textit{bimodal chirotaxis} [$\Omega_{\rm Ch}$; see Eq.~(\ref{eq:Ch})]. We find that a pusher swimmer $(\Omega_{\rm Ch}<0)$ aligns itself along the directions $\pm\mathbf{e}$ whereas a puller $(\Omega_{\rm Ch}>0)$ orbits in a plane perpendicular to $\mathbf{e}$.
Moreover, a chiral swimmer with an additional intrinsic angular velocity exhibits \textit{unimodal chirotaxis} along a single direction (see Supplemental Material for a video~\cite{SM}).\nocite{goldstein2002classical}
For symmetry reasons, chirotaxis is only possible in quadratic order or higher in odd viscosity. 
This novel orientational dynamics is therefore closely related to the exact solution of the swirling flow problems, as we demonstrate below.

\textit{Odd Stokes flows.}---We start by considering the three-dimensional (3D) incompressible force-free Stokes flow with the governing equations $\partial_j\sigma_{ij}  =0$ and $\partial_iv_i = 0$, where the stress tensor is given by $\sigma_{ij}=-p\delta_{ij}+\eta_{ijk\ell}\partial_\ell v_k$ with the velocity field $\mathbf{v}$, the pressure field $p$, and the viscosity tensor $\boldsymbol{\eta}$.
For a fluid with odd viscosity, the viscosity tensor $\boldsymbol{\eta}=\boldsymbol{\eta}^{\rm e}+\boldsymbol{\eta}^{\rm o}$ consists of a symmetric (even) part $\boldsymbol{\eta}^{\rm e}$ and an antisymmetric (odd) part $\boldsymbol{\eta}^{\rm o}$~\cite{avron1995, avron1998}, which have the following relations under the exchange of indices $ij\leftrightarrow k\ell$, 
$\eta^{\rm e}_{ijk\ell}=\eta^{\rm e}_{k\ell ij}$ and 
$\eta^{\rm o}_{ijk\ell}=-\eta^{\rm o}_{k\ell ij}$.
As a minimal model we consider a fluid with a single odd viscosity~\cite{markovich2021}.
The tensor $\boldsymbol{\eta}^{\rm o}$ can therefore be characterized by a constant coefficient, which we denote by $\eta^{\rm o}$.
In this case, the fluid breaks parity symmetry about a fixed axis while maintaining cylindrical symmetry about it~\cite{fruchart2023odd}. 
If we denote the axis by $\mathbf{e}$, such systems can be realized by fluids composed of microscopic particles collectively rotating about $\mathbf{e}$ (Fig.~\ref{fig:1})~\cite{markovich2021}. The stress tensor then reads
\begin{align}
    \sigma_{ij}
    =
    -p\delta_{ij}+2\eta^{\rm e}E_{ij}
    +
    \eta^{\rm o}
    e_\ell
    (
    \epsilon_{\ell ik}E_{kj}
    +
    \epsilon_{\ell jk}E_{ki}
    )
    ,
\label{eq:stress}
\end{align}
where $\eta^{\rm e}$ is the even (shear) viscosity, $E_{ij}=(\partial_iv_j+\partial_jv_i)/2$ is the strain-rate tensor, and $\boldsymbol{\epsilon}$ is the 3D Levi-Civita tensor.
The odd Stokes equation thus reads~\cite{markovich2021}
\begin{align}
    \eta^{\rm e}\nabla^2 \mathbf{v}-\frac{\eta^{\rm o}}{2}
    \mathbf{e}\cdot\nabla\boldsymbol{\omega}
    -\nabla P =\mathbf{0}
    ,
    \label{eq:stokes}
\end{align}
with the 
vorticity $\boldsymbol{\omega}=\nabla\times\mathbf{v}$ and the effective pressure $P=p-\eta^{\rm o}\mathbf{e}\cdot\boldsymbol{\omega}$.
Equivalently, one finds through $\nabla\cdot\mathbf{v}=0$ 
\begin{align}
  \eta^{\rm e}\nabla^2\mathbf{v}
  -
  \frac{\eta^{\rm o}}{2}
  \mathbf{e}\times
  (\nabla^2\mathbf{v})
  -\nabla P^\ast =\mathbf{0}
  ,
  \label{eq:stokes1}
\end{align}
where we have redefined the pressure $P^\ast=p-(\eta^{\rm o}/2)\mathbf{e}\cdot\boldsymbol{\omega}$.
Typically, 2D systems with odd viscosity are defined by a flat and thin fluid layer normal to the direction $\mathbf{e}$~\cite{avron1998, lapa2014, ganeshan2017, banerjee2017}.
Assuming no spatial dependence of the velocity and pressure fields along $\mathbf{e}$, the second term in Eq.~(\ref{eq:stokes}) vanishes, which recovers the classical Stokes equation with the effective pressure $P$.

In the following, we show that the 3D velocity field is unaffected by odd viscosity in the case of any swirling flow, i.e., a purely azimuthal flow resulting from the rotation of an axisymmetric body.
We start by noting that in the absence of odd viscosity, the swirling flow is pressureless, $p=\text{const}$, for symmetry reasons. In the classical Stokes equation, these flows satisfy $\nabla^2\mathbf{v}=\mathbf{0}$. 
It is evident from Eq.~(\ref{eq:stokes1}) that the same flow satisfies the odd Stokes equation with the pressure
$p=(\eta^{\rm o}/2) \mathbf{e}\cdot\boldsymbol{\omega}$. 
We have thus shown that all flows that are pressureless in conventional fluids remain unaltered in the presence of odd viscosity.
The finite pressure introduced here counterbalances the extra stress due to odd viscosity.
Since the far-field swirling flow (representing a rotlet) decays as $\mathcal{O}(r^{-2})$, any rotating axisymmetric body does not experience any force~\cite{happel2012low}, even in the presence of odd viscosity.
The torque, on the other hand, is expected to be at most linear in $\eta^{\rm o}$, because higher order dependence is forbidden, as long as the velocity does not include $\eta^{\rm o}$.
Despite its resemblance, the expression for the effective pressure $P^\ast$ differs from $P$ [defined in Eq.~(\ref{eq:stokes})] that absorbs the effects of odd viscosity in 2D fluids~\cite{ganeshan2017, banerjee2017}.
The apparent contradiction in 2D flows with $\nabla^2\mathbf{v}=\mathbf{0}$, where both prefactors should hold, is easily resolved because $\nabla\boldsymbol{\omega}$ vanishes in a 2D flow that is pressureless~\cite{Pressure}.

\textit{Rigid rotation of a sphere.}---We now solve the flow around a no-slip sphere of radius $a$ rotating with the angular velocity $\hat{\boldsymbol{\Omega}}$.
We label all quantities with the symbol $\hat{}$ because they will take the role of the auxiliary problem in the next section. 
The no-slip boundary states
$\hat{\mathbf{v}} = \hat{\boldsymbol{\Omega}}\times\mathbf{r}=a \hat{\boldsymbol{\Omega}}\times\mathbf{n}$ at $r=a$, where $\mathbf{n}$ is a surface normal pointing into the fluid. 
In classical fluids, the solution is a rotlet 
\begin{math}
\hat{\mathbf{v}}=(a^3/r^3) \hat{\boldsymbol{\Omega}}
    \times
    \mathbf{r}
\end{math}~\cite{happel2012low}.
The corresponding strain-rate tensor is then obtained as $\hat{\mathbf{E}}=3 a^3/(2r^5)[(\mathbf{r}\times\hat{\boldsymbol{\Omega}})\mathbf{r}+\mathbf{r}(\mathbf{r}\times\hat{\boldsymbol{\Omega}})]$.
Because of $\nabla^2\hat{\mathbf{v}}=\mathbf{0}$, the rotlet is pressureless, like any other swirling flow. 
As shown above, the same flow also satisfies the odd Stokes equation by setting the pressure $\hat{p}=(\hat{\eta}^{\rm o}/2) {\mathbf{e}\cdot\hat{\boldsymbol{\omega}}}$.
Since the vorticity calculated above is expressed with a source dipole~\cite{chwang_wu_1974}, the pressure takes the form
\begin{math}
    \hat{p}
    =
    -
    [\hat{\eta}^{\rm o} a^3/(2r^3)]
    \mathbf{e}\cdot
    \left(
    \mathbf{I}
    -
    3
    \mathbf{rr}/r^2
    \right)
    \cdot
    \hat{\boldsymbol{\Omega}}
    .
\end{math}
Inserting the strain-rate tensor and the pressure into Eq.~(\ref{eq:stress}) gives the full expression of the stress tensor. The traction acting on the sphere surface $\hat{\mathbf{f}}=\hat{\boldsymbol{\sigma}}\cdot\mathbf{n}$ is thus obtained as 
\begin{align}
    \hat{\mathbf{f}} 
    &=
    -3
    \hat{\eta}^{\rm 
    e}
    \bigg(
    \hat{\boldsymbol{\Omega}}\times
    \mathbf{n}
    +
    \frac{\hat{\lambda}}{2}
    (\hat{\boldsymbol{\Omega}}\times\mathbf{n})\times\mathbf{e}
    +
    \frac{\hat{\lambda}}{3}
    (\hat{\boldsymbol{\Omega}}\cdot\mathbf{e})
    \mathbf{n}
    \bigg)
    ,
    \label{eq:fns}
\end{align}
where we have defined the ratio of the odd to even viscosities as $\hat{\lambda}=\hat{\eta}^{\rm o}/\hat{\eta}^{\rm e}$.
The total force is zero, $\hat{\mathbf{F}} =\int_\mathcal{S}dS\, \hat{\mathbf{f}} =\mathbf{0}$, consistent with the discussion above, and the torque becomes
\begin{align}
    \hat{\mathbf{T}} 
    &=
    -8\pi\hat{\eta}^{\rm e}a^3
    \bigg(
    \hat{\boldsymbol{\Omega}}
    +
    \frac{\hat{\lambda}}{4}
    \hat{\boldsymbol{\Omega}}\times\mathbf{e}
    \bigg)
    .
    \label{eq:tns}
\end{align}
Equations~(\ref{eq:fns}) and (\ref{eq:tns}) are exact in fluids with odd viscosity $\hat{\lambda}$.
From the property of a pseudovector under reflection symmetry, the torque parallel to $\mathbf{e}$ remains unchanged under $\hat{\lambda}\to-\hat{\lambda}$, implying that the component $\hat{\mathbf{T}}\cdot\mathbf{e}$ must be of even order in $\hat{\lambda}$.
However, as discussed above, the torque on a rotating axisymmetric body is at most linear in $\hat{\lambda}$, and therefore the component becomes independent of the odd viscosity.
For a sphere, $\hat{\mathbf{T}}\cdot\mathbf{e}$ obeys classical Stokes' law~\cite{happel2012low}.
For the obtained torque on a sphere, the resistance tensor is $\hat{\boldsymbol{\mathcal{R}}}=8\pi\hat{\eta}^{\rm e}a^3[\mathbf{I}+(\hat{\lambda}/4)\boldsymbol{\epsilon}\cdot\mathbf{e}]$, and the energy dissipation rate is then given by $\hat{\boldsymbol{\Omega}}\cdot\hat{\boldsymbol{\mathcal{R}}}\cdot\hat{\boldsymbol{\Omega}}=8\pi\hat{\eta}^{\rm e}a^3 |\hat{\boldsymbol{\Omega}}|^2$.
The odd viscosity does not contribute to any dissipation with rigid rotation, unlike in the case of translational motion of a sphere~\cite{everts2024dissipative}.
Since the antisymmetric part of the resistance tensor is odd in $\hat{\lambda}$, reversing its sign makes the tensorial structure symmetric under $i\leftrightarrow j$, namely, $\hat{\mathcal{R}}_{ij}(\hat{\lambda})=\hat{\mathcal{R}}_{ji}(-\hat{\lambda})$.
In fact, the resistance tensors for rigid particles of any shape have the symmetry $\mathcal{R}_{ij}(\boldsymbol{\eta}^{\rm e},\boldsymbol{\eta}^{\rm o})=\mathcal{R}_{ji}(\boldsymbol{\eta}^{\rm e},-\boldsymbol{\eta}^{\rm o})$, which is known as the Onsager--Casimir reciprocity~\cite{fruchart2023odd} and follows from the generalized Lorentz reciprocal theorem~\cite{hosaka2023lorentz}, as detailed below.

\begin{figure*}[tb]
\centering
\includegraphics[width=\textwidth]{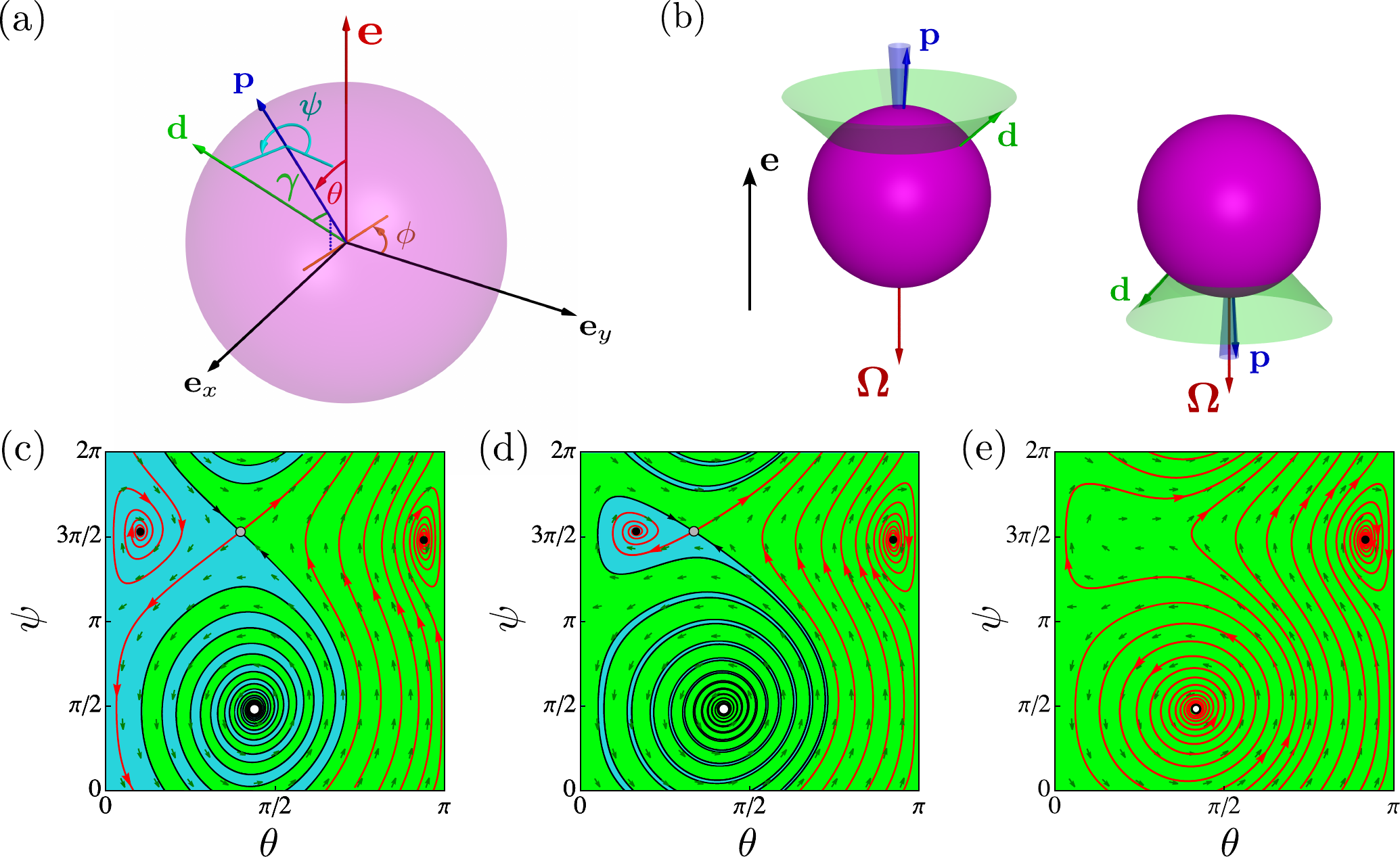}
\caption{
(a) Orientation of a chiral swimmer (magenta colored sphere) with the body-fixed axes $\mathbf{p}$ and $\mathbf{d}$ (angle $\gamma$ between them), rotated by the Euler angles $(\phi,\theta,\psi)$ within the laboratory frame $(\mathbf{e}_x,\mathbf{e}_y,\mathbf{e}_z)$ with the axis of chirality $\mathbf{e}_z\equiv\mathbf{e}$.
(b) Steady-state swimmer orientation in the two stable fixed points from panel (c). Above the bifurcation, one fixed point vanishes, and swimmers always orient towards the $-\mathbf{e}$ direction ($+\mathbf{e}$ for $\lambda<0$ or $\Omega^{\rm e}\cos\gamma<0$).
(c)--(e) Phase portraits of the orientational dynamics of a chiral pusher [Eq.~(\ref{eq:Eangles})] for different values of the intrinsic rotation rate (c) $\Omega^{\rm e}/[|\lambda B_2|/(20a)]=1$, (d) $4/3$, and (e) $5/3$, while keeping $\lambda=0.3$, $B_2<0$, and $\gamma=\pi/4$.
Below the critical value ($\Omega^{\rm e}<\Omega^{\rm e}_{\rm c}$) [panels (c) and (d)], the phase space contains a 
spiral source (white circle), a saddle point (gray circle), and two spiral sinks (black circles). Their basins of attraction are marked by blue and green regions. The red lines represent example trajectories. 
As $\Omega^{\rm e}$ is increased, the sink and the saddle approach each other, undergoing a saddle-node bifurcation at $\Omega^{\rm e}=\Omega^{\rm e}_{\rm c}$. 
For $\Omega^{\rm e}>\Omega^{\rm e}_{\rm c}$ [panel (e)], a single stable fixed point remains, characteristic of unimodal chirotaxis.
\label{fig:3}
}
\end{figure*}%

\textit{Microswimmer's angular velocity.}---We can now use the flow solution around a rotating sphere and the Lorentz reciprocal theorem to determine the angular velocity of a spherical microswimmer.
The Lorentz reciprocal theorem provides an integral identity connecting the velocity and stress of two flow problems~\cite{masoud2019}.
It provides an elegant way to determine the swimming velocity of an active swimmer without explicitly solving its flow. Although odd viscosity violates the reciprocity, we have previously demonstrated that the Lorentz reciprocal theorem can still be applied if the odd viscosity has the opposite sign between the main and the auxiliary problems, $\hat\eta^{\rm o}=- \eta^{\rm o}$, while keeping the same sign for the even viscosities, $\hat\eta^{\rm e}=\eta^{\rm e}$~\cite{hosaka2023lorentz}.
We consider the surface-driven microswimmer with a prescribed effective slip velocity $\mathbf{v}^{\rm s}$ in a fluid characterized by $\lambda=\eta^{\rm o}/\eta^{\rm e}$ as the main and the rotating sphere with surface velocity $\hat{\mathbf{v}}$ as the auxiliary problem.
We consider the rapid relaxation of the rotational axis of the fluid components in the direction $\mathbf{e}$, which leads to the constant torque density due to the rotating constituents.
This ensures that the odd viscosity remains constant in both space and time~\cite{markovich2021}, as previously assumed~\cite{khain2022, hosaka2023lorentz, everts2024dissipative}.
The reciprocal theorem can be written in the form
\begin{math}
    \boldsymbol{\Omega}\cdot\hat{\mathbf{T}} 
    =
    -\int_\mathcal{S}dS\, \mathbf{v}^{\rm s}\cdot\hat{\mathbf{f}} 
\end{math}~\cite{stone1996propulsion}, where $\boldsymbol{\Omega}$ is the angular velocity of a torque-free swimmer $(\mathbf{T}=\mathbf{0})$, $\hat{\mathbf{T}}$ is the torque on the sphere given by Eq.~(\ref{eq:tns}), and $\hat{\mathbf{f}}$ is its surface traction (\ref{eq:fns}). The theorem holds for any angular velocity vector of the auxiliary problem $\hat{\boldsymbol{\Omega}}$ and therefore uniquely determines $\boldsymbol{\Omega}$.
We can analytically obtain $\boldsymbol{\Omega}$, which comprises contributions from the even and odd viscosities $\boldsymbol{\Omega}=\boldsymbol{\Omega}^{\rm e}+\boldsymbol{\Omega}^{\rm o}$
(see Appendix~A for details).
The even angular velocity $\boldsymbol{\Omega}^{\rm e}=-3/(8\pi a^3)\int_\mathcal{S}dS\,\mathbf{n}\times\mathbf{v}^{\rm s}$ describes the rotation rate of the slip velocity~\cite{stone1996propulsion}, while the odd part is given by
\begin{align}
    \boldsymbol{\Omega}^{\rm o}
    =
    \frac{3\lambda}{10a^3}
    \mathbf{e}\cdot\mathbf{S}
    \cdot
    \left[
    \mathbf{I}
    +
    g_1(\lambda)
    \boldsymbol{\epsilon}\cdot\mathbf{e}
    +
    g_2(\lambda)
    (\mathbf{ee}-\mathbf{I})
    \right]
    .
    \label{eq:Omegao}
\end{align}
Here, $\mathbf{S}=\int_\mathcal{S}dS\,
\left[-(5/2)(\mathbf{v}^{\rm s}\mathbf{n}+\mathbf{n}\mathbf{v}^{\rm s})+(\mathbf{n}\cdot\mathbf{v}^{\rm s})\mathbf{I}\right]/(8\pi)$ is the stresslet~\footnote{The stresslet defined as $\mathbf{S}$ with a prefactor $8\pi\eta^{\rm e}$ is equivalent to Eq.~(64) in Ref.~\cite{nasouri2018higher}}\nocite{nasouri2018higher}, $g_n(\lambda)\equiv(\lambda/4)^n/[1+(\lambda/4)^2]$, and we have assumed volume conservation within the swimmer $(\int_\mathcal{S}dS\, \mathbf{n}\cdot\mathbf{v}^{\rm s}=0)$.
To the first order in $\lambda$, only the stresslet along the direction $\mathbf{e}$ determines the rotation $\boldsymbol{\Omega}^{\rm o}$, while higher order terms contain contributions transverse to $\mathbf{e}\cdot\mathbf{S}$.
Since the rotation vector $\boldsymbol{\Omega}^{\rm e}$ and the stresslet $\mathbf{S}$ represent distinct moments, $\boldsymbol{\Omega}^{\rm e}$ and $\boldsymbol{\Omega}^{\rm o}$ do not couple with each other for a given slip velocity $\mathbf{v}^{\rm s}$.

In 2D incompressible flows, the flow field and effective pressure are independent of odd viscosity as long as the boundary conditions include only the velocity field (e.g., no-slip boundaries)~\cite{ganeshan2017}. In this case, the force on any object is unaffected by the odd viscosity, and the torque it generates vanishes only for area-conserving bodies~\cite{lapa2014,ganeshan2017}. These results indicate that a 2D swimmer of any shape, with constant area and boundary conditions involving only the flow field, does not exhibit any extra translational or angular velocity due to odd viscosity (see also Ref.~\cite{lapa2014} for a different approach).
The rotational dynamics, described by Eq.~(\ref{eq:Omegao}), is therefore characteristic of 3D odd flows.

\textit{Bimodal chirotaxis.}---We first study the dynamics of an achiral microswimmer without intrinsic rotation, which requires $\boldsymbol{\Omega}^\mathrm{e}=\mathbf{0}$ due to symmetry.
By assuming a general axisymmetric profile, the stresslet is given by a force dipole
$\mathbf{S}=(a^2B_2/6)(3\mathbf{pp}-\mathbf{I})$ with the preferred direction $\mathbf{p}$~\cite{Ishikawa.Pedley2006, lauga2020fluid}.
From this, we arrive at Eq.~(\ref{eq:intro}) with the angular velocities
\begin{align}
\Omega_{\rm Sp}&=
\frac{3g_1(\lambda)B_2}{5a},
\label{eq:Sp} \\
\Omega_{\rm Pr}&=-
\frac{\lambda B_2}{20a}[1-3g_2(\lambda)(\mathbf{e}\cdot\mathbf{p})^2], 
\label{eq:Pr}\\
\Omega_{\rm Ch}&=
\frac{3g_2(\lambda)B_2}{5a}.
\label{eq:Ch}
\end{align}
The sign of the parameter $B_2$ determines the type of microswimmers: $B_2>0$ corresponds to pullers, which generate thrust from the front, while $B_2<0$ corresponds to pushers, which generate thrust from the back~\cite{lauga2020fluid}.
With $\dot{\mathbf{p}}=\boldsymbol{\Omega}^{\rm o}\times\mathbf{p}$, its orientational dynamics is determined by
\begin{align}
    \dot{\mathbf{p}}
    =
    \Omega_{\rm Pr}
    (\mathbf{e}\times\mathbf{p})
    +
    \Omega_{\rm Ch}
    (\mathbf{e}\cdot\mathbf{p})
    (\mathbf{e}\times\mathbf{p})\times \mathbf{p}
    \,.
    \label{eq:pdot}
\end{align}
Figure~\ref{fig:2}(a) shows example trajectories of the orientation vector $\mathbf{p}$ following from this equation (a closed-form solution is derived in Appendix~B).
The first term in Eq.~(\ref{eq:pdot}) describes the precession about the axis $\mathbf{e}$ and the second represents the relaxation dynamics towards or away from this axis.
In particular, pullers reorient perpendicular to the direction $\mathbf{e}$ (cyan line), while pushers orient up or down (magenta line), depending on the initial orientation.
The leading alignment effects are of $\mathcal{O}(\lambda^2)$ with a relaxation rate $3\lambda^2|B_2|/(80a)$, while the precession is $\mathcal{O}(\lambda)$ with a frequency $|\lambda B_2|/(20a)$.
The bimodal alignment closely resembles densitaxis~\cite{shaik2024densitaxis} and similar behavior in nematic liquid crystals~\cite{lintuvuori2017hydrodynamics,chi2020Surface}.
However, unlike these cases, chirotaxis is driven purely by the fluid chirality, without spatial inhomogeneity or elastic torque.

Expression~(\ref{eq:pdot}) is similar to the Landau--Lifshitz--Gilbert equation for spin systems~\cite{lakshmanan2011fascinating} except for the additional factor $({\mathbf{e}\cdot\mathbf{p}})$ in the second term, also known as the damping term.
This difference can be understood from the fundamental symmetry properties. Since the orientation $\mathbf{p}$ is a polar vector and $\mathbf{e}$ a pseudovector, they cannot show alignment in a single direction; if they can align parallel, they can also align antiparallel.
From this symmetry, the alignment terms in Eq.~(\ref{eq:pdot}) are of an even order in $\lambda$.
On the other hand, magnetization or spin is a pseudovector and always aligns parallel to a magnetic field, also a pseudovector.
Although the symmetry arguments above allow a unidirectional alignment for a chiral swimmer, the question whether it actually takes place requires an analysis of the orientation dynamics, as explored in the dynamical system described in the next section.

A swimmer moving in the direction $\mathbf{p}$ with a translational velocity determined by the squirming mode $B_1$, $\mathbf{V}=(2/3)B_1\mathbf{p}$, can be characterized by the squirming parameter $\beta=B_2/B_1$~\cite{Ishikawa.Pedley2006}.
Note that the velocity $\mathbf{V}$ is independent of $\lambda$, as derived in Appendix~C.
Example trajectories in Fig.~\ref{fig:2}(b) show that a puller tends to align with the plane perpendicular to the $\mathbf{e}$-axis where it eventually moves on circular paths with radius $d=(40a/3)|\beta\lambda|$.
Although such circular dynamics are known for a dipolar microswimmer in a 2D \textit{compressible} flow with odd viscosity~\cite{hosaka2023hydrodynamics}, alignment behavior is not possible in 2D, and thus chirotaxis is unique in 3D systems.
A pusher, on the other hand, approaches in the direction  $\pm\mathbf{e}$, accompanied by a transient helical motion with pitch $h=(80\pi a/3)/|\beta\lambda[1-3g_2(\lambda)]|$.

\textit{Unimodal chirotaxis.}---We now investigate the chirotactic behavior of a chiral swimmer that has, in addition to the force dipole, a nonvanishing intrinsic angular velocity $\boldsymbol{\Omega}^{\rm e}=\Omega^{\rm e}\mathbf{d}$ [Fig.~\ref{fig:1}(b)]. 
The direction $\mathbf{d}$ is fixed in the body frame and forms an angle $\gamma$ with the axis $\mathbf{p}$.
We describe the swimmer rotation with the Euler angles $(\phi,\theta,\psi)$~\cite{goldstein2002classical}, 
where $\phi$ and $\theta$ are the azimuthal and polar angles measured from the $\mathbf{e}$-axis and $\psi$ is the roll angle about the direction $\mathbf{p}$ [Fig.~\ref{fig:3}(a)].
The equations of motion
then read (see Appendix~D for a derivation)
\begin{align}
    \dot{\phi}
    &=
    \Omega_{\rm Pr}
    +
    \Omega^{\rm e}\,
    \frac{\sin\gamma\sin\psi}{\sin\theta},
    \nonumber\\
    \dot{\theta}
    &=
    \Omega_{\rm Ch}
    \sin\theta\cos\theta
    +
    \Omega^{\rm e}
    \sin\gamma\cos\psi
    ,
    \label{eq:Eangles}\\
    \dot{\psi}
    &=
    \Omega_{\rm Sp}\cos\theta
    +
    \Omega^{\rm e}
    \left(
    \cos\gamma
    -\frac{\sin\gamma\sin\psi}{\tan\theta}
    \right)
    ,
    \nonumber
\end{align}
where $\Omega_{\rm Pr}$ is evaluated by setting $\mathbf{e}\cdot\mathbf{p}=\cos\theta$ in Eq.~(\ref{eq:Pr}).
The equations do not depend on $\phi$, so it is sufficient to study the phase portraits in the $(\theta,\psi)$ plane [Fig.~\ref{fig:3}(c)--(e)]. 
Since the vector $(\dot\theta,\dot\psi)$ describes a continuous field on a sphere, the flow has to satisfy the Poincar\'{e}--Hopf index theorem stating that Euler characteristic (the number of sources and sinks reduced by the number of saddle points) gives 2~\cite{perko2001}.
At low angular velocities $\Omega^{\rm e}$ [Fig.~\ref{fig:3}(c)], the system has a spiral source and a saddle point, as well as two spiral sinks that are characteristic of bimodal chirotaxis.
When $\Omega^{\rm e}$ is increased, one of the sinks annihilates with the saddle at a saddle-node bifurcation~\cite{strogatz}. Above the bifurcation, a single stable fixed point remains, indicating the emergence of unimodal chirotaxis [Fig.~\ref{fig:3}(e)].
The critical value is, to the leading order, $\Omega^{\rm e}_{\rm c}/[|\lambda B_2|/(20a)]=3(\sin^{2/3}\gamma+\cos ^{2/3}\gamma)^{-3/2}$~\cite{SM}. The lowest angular frequency needed for unimodal chirotaxis is therefore achieved when the angle between intrinsic rotation axis $\mathbf{d}$ and the swimmer axis $\mathbf{p}$ is $\gamma=\pi/4$. Unimodal chirotaxis is possible because the chiral swimmer is characterized by a pseudovector $\mathbf{d}$, which can align itself with the pseudovector $\mathbf{e}$, describing the axis of odd viscosity.

In conclusion, we have presented exact results that allowed us to uncover alignment response of a microswimmer in a fluid with odd viscosity. Our theoretical findings should be applicable to a variety of transport phenomena in chiral active fluids.
In particular, examples of living systems range from bacterial suspensions to actomyosin or active chiral biopolymer networks in which torques are generated internally~\cite{markovich2021}. 
Our analytical solution for a swirling flow could further be applied to the collective behavior of rotating constituents in synthetic and living chiral active systems~\cite{soni2019odd, massana2021arrested, bililign2022motile, drescher2009dancing, tan2022odd}.

\acknowledgments
We acknowledge support from the Max Planck Center Twente for Complex Fluid Dynamics, the Max Planck Society, the Max Planck School Matter to Life, and the MaxSynBio Consortium, which are jointly funded by the Federal Ministry of Education and Research (BMBF) of Germany.
A.V.\ acknowledges support from the Slovenian Research Agency (Grant No.\ P1-0099).
\\
\appendix

\begin{widetext}
\setcounter{equation}{0}
\renewcommand\theequation{A\arabic{equation}}

\textit{Appendix~A: Derivation of the angular velocity $\boldsymbol{\Omega}$.---}We derive the general expression for the angular velocity of a spherical swimmer of radius $a$ in fluids with the odd viscosity $\lambda=\eta^{\rm o}/\eta^{\rm e}$.
By inserting the traction $\hat{\mathbf{f}}$ (\ref{eq:fns}) and the torque $\hat{\mathbf{T}}$ (\ref{eq:tns}) on a rigidly rotating sphere into the Lorentz reciprocal theorem, we find
\begin{align}
    \boldsymbol{\Omega}\cdot
    \left(
    \mathbf{I}
    -
    \frac{\lambda}{4}
    \boldsymbol{\epsilon}\cdot\mathbf{e}
    \right)
    \cdot
    \hat{\boldsymbol{\Omega}}
    =
    -
    \frac{3}{8\pi a^3}
    \int_\mathcal{S} dS\,
    \mathbf{v}^{\rm s}
    \cdot
    \left(
    \hat{\boldsymbol{\Omega}}\times\mathbf{n}
    -
    \frac{\lambda}{2}
    (\hat{\boldsymbol{\Omega}}\times\mathbf{n})\times\mathbf{e}
    -
    \frac{\lambda}{3}
    (\hat{\boldsymbol{\Omega}}\cdot\mathbf{e})\mathbf{n}
    \right)
    ,
\end{align}
where we have used the relations between the main and the auxiliary problems, $\hat{\eta}^{\rm e}=\eta^{\rm e}$ and $\hat{\eta}^{\rm o}=-\eta^{\rm o}$ (or equivalently, $\hat{\lambda}=-\lambda$)~\cite{hosaka2023lorentz}. This equation is valid for any angular velocity $\hat{\boldsymbol{\Omega}}$, which can therefore 
be left out from both sides of the above equation:
\begin{align}
    \boldsymbol{\Omega}\cdot
    \left(
    \mathbf{I}
    -
    \frac{\lambda}{4}
    \boldsymbol{\epsilon}\cdot\mathbf{e}
    \right)
    =
    -
    \frac{3}{8\pi a^3}
    \int_\mathcal{S} dS\,
    \left[
    (\mathbf{n}\times\mathbf{v}^{\rm s})
    \cdot
    \left(
    \mathbf{I}
    -
    \frac{\lambda}{4}
    \boldsymbol{\epsilon}\cdot\mathbf{e}
    \right)
    +
    \frac{\lambda}{4}
    \mathbf{e}\cdot
    (\mathbf{v}^{\rm s}\mathbf{n}
    +
    \mathbf{n}\mathbf{v}^{\rm s})
    -
    \frac{5\lambda}{6}
    (\mathbf{v}^{\rm s}\cdot\mathbf{n})\mathbf{e}
    \right]
    .
\end{align}
From the inverse of the tensor $[\mathbf{I}-(\lambda/4)\boldsymbol{\epsilon}\cdot\mathbf{e}]^{-1}=[\mathbf{I}+g_1(\lambda)\boldsymbol{\epsilon}\cdot\mathbf{e}+g_2(\lambda)(\mathbf{ee}-\mathbf{I})]$ with $g_n(\lambda)\equiv(\lambda/4)^n/[1+(\lambda/4)^2]$, we obtain the solution for the angular velocity of the swimmer
\begin{align}
    \boldsymbol{\Omega}
    =
    -
    \frac{3}{8\pi a^3}
    \int_\mathcal{S} dS\,
    \left(
    \mathbf{n}\times\mathbf{v}^{\rm s}
    +
    \frac{\lambda}{4}
    \mathbf{e}\cdot
    (\mathbf{v}^{\rm s}\mathbf{n}
    +
    \mathbf{n}\mathbf{v}^{\rm s})
    \cdot
    [\mathbf{I}+g_1(\lambda)\boldsymbol{\epsilon}\cdot\mathbf{e}+g_2(\lambda)(\mathbf{ee}-\mathbf{I})]
    -
    \frac{5\lambda}{6}
    (\mathbf{v}^{\rm s}\cdot\mathbf{n})\mathbf{e}
    \right)
    .
\end{align}
From the volume conservation of flows within the swimmer $(\int_\mathcal{S}dS\, \mathbf{n}\cdot\mathbf{v}^{\rm s}=0)$, we can express the angular velocity in terms of the stresslet $\mathbf{S}$ as in Eq.~(\ref{eq:Omegao}).

\setcounter{equation}{0}
\renewcommand\theequation{B\arabic{equation}}
\textit{Appendix~B: Solution of the equations of motion of an achiral swimmer.---}We derive the equations of motion for the orientation $\mathbf{p}$ of a microswimmer. 
From Eq.~(\ref{eq:pdot}), the angular dynamics is determined by
\begin{equation}
\begin{aligned}
     \dot{\theta} =
     \frac{3g_2(\lambda)B_2}{10a} \sin(2\theta),
     \quad
     \dot{\phi} =
     -
     \frac{\lambda B_2}{20a}
    [1-3g_2(\lambda)\cos^2\theta]
    ,
    \label{eq:thetatphit}
\end{aligned}
\end{equation}
where $\theta$ and $\phi$ are the polar and azimuthal angles of the vector $\mathbf{p}$ in spherical coordinates when the $\mathbf{e}$-direction is chosen as a positive polar axis.
The angular dynamics is solely determined by $\theta$ as the system has cylindrical symmetry about $\mathbf{e}$.
Setting the initial azimuthal angle as $\phi_0=0$ without loss of generality, the equations are solved by
\begin{align}
    \tan\theta(t) =
    \tan\theta_0
    \exp\left[\frac{3g_2(\lambda)B_2}{5a}t\right]
    ,
    \quad
    \phi(t)
    =
    -
    \frac{\lambda B_2}{20a}t
    +
    \frac{\lambda}{4}
    \ln
    \left(
    \frac{\sin\theta(t)}{\sin\theta_0}
    \right)
    .
\end{align}

\setcounter{equation}{0}
\renewcommand\theequation{C\arabic{equation}}
\textit{Appendix~C: Translational velocity of a swimmer with the squirming mode $B_1$.---}We show that a neutral spherical squirmer with the lowest surface mode $B_1$ moves with velocity $\mathbf{V}=(2/3)B_1\mathbf{p}$, which is independent of the odd viscosity. In classical fluids without $\eta^{\rm o}$, by imposing the effective slip velocity $\mathbf{v}^{\rm s}=-B_1(\mathbf{I}-\mathbf{nn})\cdot\mathbf{p}$ on the swimmer surface, the resulting flow field in the laboratory frame is expressed with a source dipole $\mathbf{v}=-(a^3B_1/3) (\mathbf{I}/r^3-3\mathbf{rr}/r^5)\cdot\mathbf{p}$~\cite{lauga2020fluid}.
Since the source dipole is irrotational and is not associated with any pressure, the same pressureless flow also satisfies the odd Stokes equation~(\ref{eq:stokes1}), as discussed in the main text. At the same time, the swimmer in the identical flow field remains force- and torque-free in the presence of $\eta^{\rm o}$. This can be seen by considering the far-field limit of the stress tensor, which is of order $\mathcal{O}(r^{-4})$, and therefore indicates that the net hydrodynamic force and torque on the particle is zero~\cite{happel2012low}.
The swimming velocity is therefore not affected by any value of the odd viscosity, extending the result previously derived to linear order only~\cite{hosaka2023lorentz}.
In general, the direction of self-propulsion can be different from that of the force dipole $\mathbf{p}$. The steady-state motion of such a nonaxisymmetric swimmer is helical along the axis of chirality $\mathbf{e}$ or along a circle in the plane perpendicular to $\mathbf{e}$.
This is in contrast to the case $\mathbf{V}\parallel\mathbf{p}$ studied in the main text, where the stationary trajectory converges to a straight line or a circle.

\setcounter{equation}{0}
\renewcommand\theequation{D\arabic{equation}}
\textit{Appendix~D: Derivation of Eq.~(\ref{eq:Eangles}).---}To derive the orientational dynamics of the chiral swimmer whose surface velocity has, in addition to the force dipole, a nonvanishing intrinsic angular velocity $\Omega^{\rm e}\mathbf{d}$, we express the angular velocity as 
\begin{equation}
    \boldsymbol{\Omega}
    =
    \boldsymbol{\Omega}^{\rm o}+\boldsymbol{\Omega}^{\rm e}
    = 
\Omega_{\rm Sp}(\mathbf{e}\cdot\mathbf{p})\mathbf{p} 
+ \Omega_{\rm Pr}\mathbf{e}
+ \Omega_{\rm Ch} (\mathbf{e}\cdot\mathbf{p})\mathbf{e}\times\mathbf{p} + \Omega^{\rm e}\mathbf{d}
\,.
\end{equation}
We introduce the swimmer-fixed basis $(\mathbf{e}_1,\mathbf{e}_2,\mathbf{e}_3)$ with $\mathbf{e}_3\equiv \mathbf{p}$, which is rotated from the laboratory frame by the Euler angles $(\phi,\theta,\psi)$ (see Supplemental Material for details~\cite{SM}).  We choose $\mathbf{e}_1$ such that the vector $\mathbf{d}$ has the direction $\mathbf{d}=\sin\gamma\mathbf{e}_1
+    \cos\gamma\mathbf{e}_3$. The odd viscosity axis $\mathbf{e}$ can be expressed in the swimmer frame with the inverse rotation
$
    \mathbf{e} = \sin\theta\sin\psi \mathbf{e}_1 + \sin\theta\cos\psi  \mathbf{e}_2+\cos\theta \mathbf{e}_3
$. The angular velocity in the body-fixed basis then reads
\begin{align}
    \boldsymbol{\Omega}
    =
    \Omega_{\rm Sp}\cos\theta\mathbf{e}_3
    +
    \Omega_{\rm Pr}
    [
    \sin\theta
    (\sin\psi\mathbf{e}_1
    +\cos\psi\mathbf{e}_2)
    +
    \cos\theta\mathbf{e}_3
    ]
    +
    \Omega_{\rm Ch}\sin\theta\cos\theta
    (\cos\psi\mathbf{e}_1
    -\sin\psi\mathbf{e}_2)
    +
    \Omega^{\rm e}(\sin\gamma\mathbf{e}_1
    +
    \cos\gamma\mathbf{e}_3).
    \label{eq:omegaeuler}
\end{align}
At the same time, angular velocity can be expressed with the time derivatives of the Euler angles as $\boldsymbol{\Omega}
    =
    (\dot \phi \sin\theta \sin\psi + \dot \theta \cos\psi)\mathbf{e}_1
    +
    (\dot\phi \sin\theta \cos\psi - \dot \theta \sin\psi)\mathbf{e}_2
    +
    (\dot \phi \cos\theta + \dot \psi)\mathbf{e}_3
$~\cite{goldstein2002classical}. From the equivalence of the two expressions, we obtain the equations of motion~(\ref{eq:Eangles}).
When $\Omega^{\rm e}=0$, we recover Eq.~(\ref{eq:thetatphit}) for an achiral swimmer.
\end{widetext}

\bibliography{myref}

\begin{thebibliography}{87}
\bibitem [{Pre()}]{Pressure}%
  \BibitemOpen
  \href@noop {} {\bibinfo {title} {{ In a classical pressureless Stokes flow
  ($p=\mathrm{const}$ and thus $\nabla^2\mathbf{v}=\mathbf{0}$ without
  $\eta^{\rm o}$), the curl of the vorticity is zero,
  $\nabla\times\boldsymbol{\omega}=\mathbf{0}$. Since the vorticity in 2D
  fluids has only the $z$-component
  $(\boldsymbol{\omega}=\omega_z\mathbf{e}_z)$ with
  $\mathbf{e}_z\equiv\mathbf{e}$, one finds $\partial_x\omega_z=0$ and
  $\partial_y\omega_z=0$, hence $\nabla\omega_z=\mathbf{0}$ }}}\BibitemShut
  {NoStop}%
\end{thebibliography}

\end{document}